\documentclass[letterpaper,english,reprint,aps,prl,superscriptaddress,showpacs,longbibliography]{revtex4-1}
\usepackage[latin9]{inputenc}
\usepackage{float}
\usepackage{textcomp}
\usepackage{amsmath}
\usepackage{amssymb}
\usepackage{graphicx}

\makeatletter

\pdfpageheight\paperheight
\pdfpagewidth\paperwidth

\newcommand{\lyxdot}{.}

\usepackage{babel}
\usepackage[colorlinks,
            linkcolor=red,
            anchorcolor=black,
            citecolor=blue
            ]{hyperref}

\makeatother

\usepackage{babel}
\begin{document}
\title{Broad Spiral-Bandwidth of Orbital Angular Momentum Interface between
Photon and Memory}
\author{Dong-Sheng Ding}
\email{dds@ustc.edu.cn}

\affiliation{Key Laboratory of Quantum Information, University of Science and Technology
of China, Hefei, Anhui 230026, China.}
\affiliation{Synergetic Innovation Center of Quantum Information and Quantum Physics,
University of Science and Technology of China, Hefei, Anhui 230026,
China.}
\author{Ming-Xin Dong}
\affiliation{Key Laboratory of Quantum Information, University of Science and Technology
of China, Hefei, Anhui 230026, China.}
\affiliation{Synergetic Innovation Center of Quantum Information and Quantum Physics,
University of Science and Technology of China, Hefei, Anhui 230026,
China.}
\author{Wei Zhang}
\affiliation{Key Laboratory of Quantum Information, University of Science and Technology
of China, Hefei, Anhui 230026, China.}
\affiliation{Synergetic Innovation Center of Quantum Information and Quantum Physics,
University of Science and Technology of China, Hefei, Anhui 230026,
China.}
\author{Shuai Shi}
\affiliation{Key Laboratory of Quantum Information, University of Science and Technology
of China, Hefei, Anhui 230026, China.}
\affiliation{Synergetic Innovation Center of Quantum Information and Quantum Physics,
University of Science and Technology of China, Hefei, Anhui 230026,
China.}
\author{Yi-Chen Yu}
\affiliation{Key Laboratory of Quantum Information, University of Science and Technology
of China, Hefei, Anhui 230026, China.}
\affiliation{Synergetic Innovation Center of Quantum Information and Quantum Physics,
University of Science and Technology of China, Hefei, Anhui 230026,
China.}
\author{Ying-Hao Ye}
\affiliation{Key Laboratory of Quantum Information, University of Science and Technology
of China, Hefei, Anhui 230026, China.}
\affiliation{Synergetic Innovation Center of Quantum Information and Quantum Physics,
University of Science and Technology of China, Hefei, Anhui 230026,
China.}
\author{Guang-Can Guo}
\affiliation{Key Laboratory of Quantum Information, University of Science and Technology
of China, Hefei, Anhui 230026, China.}
\affiliation{Synergetic Innovation Center of Quantum Information and Quantum Physics,
University of Science and Technology of China, Hefei, Anhui 230026,
China.}
\author{Bao-Sen Shi}
\email{drshi@ustc.edu.cn}

\affiliation{Key Laboratory of Quantum Information, University of Science and Technology
of China, Hefei, Anhui 230026, China.}
\affiliation{Synergetic Innovation Center of Quantum Information and Quantum Physics,
University of Science and Technology of China, Hefei, Anhui 230026,
China.}
\date{\today}
\begin{abstract}
The complex interactions between orbital angular momentum (OAM) light
and atoms are particularly intriguing in the areas of quantum optics
and quantum information science. Building a versatile high-dimensional
quantum network needs broad spiral-bandwidth for preparing higher-quanta
OAM mode and resolving the bandwidth mismatch in spatial space and
etc. Here, we experimentally demonstrate a broad spiral-bandwidth
quantum interface between photon and memory. Through twisted fields
of the writing and reading, the correlated OAM distribution between
photon and memory is significantly broadened. This broad spiral-bandwidth
quantum interface could be spanned in multiplexing regime and could
work in high-quanta scenario with capability of $|l|=30$, and we
demonstrate the entanglement within 2-D subspace with a fidelity of
$80.5\text{\textpm}4.8\%$ for high $l$. Such state-of-the-art technology
to freely control the spatial distribution of OAM memory is very helpful
to construct high-dimensional quantum networks and provides a benchmark
in the field of actively developing methods to engineer OAM single
photon from matters.
\end{abstract}
\maketitle
The interaction between orbital angular momentum (OAM) of structured
light and matters has many intriguing applications \citep{padgett2017orbital},
including trapping of particles \citep{he1995direct,he2009rotating}
and measuring rotation angular \citep{courtial1998measurement,lavery2013detection},
OAM-based imaging \citep{furhapter2005spiral} and optical communications
\citep{wang2012terabit}. In quantum information field, light carried
with OAM could significantly enhance the information capacity, thus
advancing the developments of the high-dimensional (high-D) quantum
networks, especially in OAM entanglement generation \citep{mair2001entanglement,dada2011experimental,fickler2012quantum,krenn2014generation},
OAM-based quantum memory \citep{inoue2009measuring,ding2013single,nicolas2014quantum,zhou2015quantum,ding2015quantum,ding2016high,zhang2016experimental}
and OAM-based teleportation \citep{wang2015quantum}. One of strumbling
block of constructing a high-D quantum network is how to establish
a versatile high-D OAM quantum interface between photon and memory
\citep{krenn2017orbital}.

Building a high-D OAM quantum interface could be based on the protocol
of Duan-Lukin-Cirac-Zoller (DLCZ) \citep{duan2001long} where the
probabilistically generated OAM photon is entangled with memory \citep{inoue2006entanglement,inoue2009measuring}.
There are many parameters to characterize the performance of the interface
between photon and memory \citep{bussieres2013prospective,heshami2016quantum},
such as lifetime, efficiency and fidelity and etc \citep{heshami2016quantum,ma2017optical,brennen2015focus}.
The most unique parameter of high-D OAM quantum interface could be
spiral-bandwidth $\delta l$, which characterizes the mode-matching
bandwidth window \citep{ding2017broad}. The adjacent nodes in high-dimensional
quantum networks may be diverse and different in spatial mode, spiral-bandwidth
and etc, for example, one is encoded in $\pm l$ OAM spaces and the
other one is in $\pm(l+m)$ OAM spaces; or one has the spatial-bandwidth
of $\delta l$ and another is $\delta(l+m)$, thus needing a technology
to make the quantum interface be more flexible and controllable and
then people can connect them freely \citep{kimble2008quantum}.

In this letter, we experimentally demonstrate a high-D OAM interface
between photon and memory in delayed spontaneously four-wave mixing
process. In this configuration, the write- and read-laser beams are
individually encoded, thus making the joint of correlation against
$l$ modes broadened because the interaction length is increased in
transverse azimuthal direction. This offers the ability to control
the spatial distribution, including entangled OAM eigenmode $\pm l$
and the spiral-bandwidth $\delta l$. Based on that, we demonstrate
the potential applications for OAM multiplexing, and give an obvious
contrast data with inputting $\Delta l=10$. We also have achieved
high-D entanglement with $l$ up to 16 and high-quanta 2-dimensional
OAM entanglement with $l$ up to 30, all of which obey the entanglement
properties. The reported results are useful for realizing broad spiral-bandwidth
and high-D quantum memory and increasing the capacity of quantum communication,
and also is a benchmark of searching ways to explore versatile quantum
interfaces.

\begin{figure}[t]
\includegraphics[width=1\columnwidth]{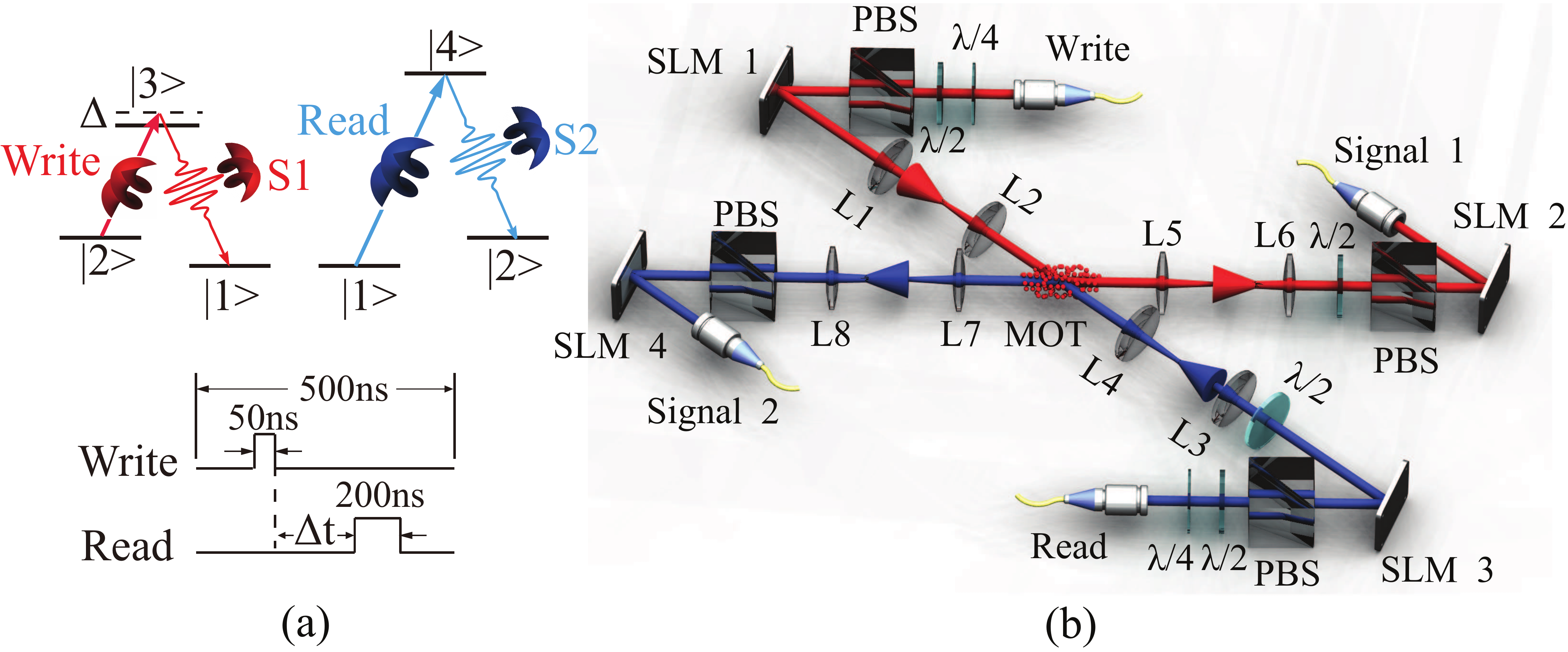}\caption{Overview of experiment. (a) The relevant energy level diagram. Write
and Read represent write-laser pulse and read-laser pulse respectively;
S1 and S2 are Signal 1 photon and Signal 2 photon. States $|1\rangle$,
$|2\rangle$, $|3\rangle$ and $|4\rangle$ correspond to $^{85}Rb$
atomic levels of $5S_{1/2}(F=2)$, $5S_{1/2}(F=3)$, $5P_{1/2}(F=3)$
and $5P_{3/2}(F=3)$ respectively. $\Delta$ represents the detuning
of write-laser pulse, which is set to be $+70\times2\pi$ MHz. The
part in below is the time sequence of experiment. (b) Experimental
setup. MOT is magneto-optical trap, $L1-L8$ represent a series of
lenses, $\lambda/4$ is quarter-wave plate, $\lambda/2$ is half-wave
plate and PBS is polarization beam splitter. SLM $1\sim4$ are spatial
light modulators.}

\label{setup}
\end{figure}
The experimental media is an optically thick atomic ensemble of Rubidium
85 ($^{85}Rb$) that is trapped in two-dimensional magneto-optic trap
(MOT). The involved schematic of the energy levels and the experimental
setup are shown in Fig. \ref{setup}(a) and Fig. \ref{setup}(b).
We firstly establish the correlation between a collective spin excited
state (spin wave, also called as atomic memory) and a single photon
(Signal 1) through spontaneous Raman scattering (SRS) in atomic ensemble.
In this process, the write-laser is set to blue-detuned with atomic
transition $|2\rangle\rightarrow|3\rangle$. After reflecting from
SLM 1 as depicted in Fig. \ref{setup} (b), the write-laser has carried
on the OAM phase message loaded by a computer. Then, a 4-f image system
with unequal arms, which is consisted of two lenses L1 and L2 with
focal length of 300 mm and 500 mm respectively, is utilized to map
the OAM phase of the write-laser to the center of atomic ensemble
accurately. The Signal 1 photon emitted from atomic ensemble is mapped
onto another SLM 2 for detecting the OAM modes. Due to the angular
momentum is conserved in SRS process, hence the spatial modes of the
spin wave and Signal 1 are entangled in OAM degree of freedom. This
OAM correlation can be flexibly demonstrated by mapping and checking
the OAM modes on SLM 1 and SLM 2 respectively.

The OAM-based DLCZ quantum memory is built when the entanglement between
the spin wave and Signal 1 photon is created. After a storage time
of $\Delta t$, we use another SLM 3 to load OAM structured light
to read the spin wave out to Signal 2, the Signal 2 is also mapped
onto another SLM 4. Ultimately, in order to check the quantum correlation
between Signal 1 and atomic spin wave, we measure the coincidence
counts between Signal 1 and Signal 2 by projecting them onto SLM 2
and SLM 4 respectively, in which the different phase structures on
both of SLM 2 and 4 are loaded for measurement. Here, two couples
of 4-f systems with unequal arms are used to map the OAM phase of
signal photons to SLM accurately, see supplementary information. The
reflected photons from SLMs are collected into two single-mode fibers,
which are detected by two detectors (avalanche diode, PerkinElmer
SPCM-AQR-16-FC, 60\% efficiency, maximum dark count rate of 25/s)
respectively.

\begin{figure}[t]
\includegraphics[width=0.8\columnwidth]{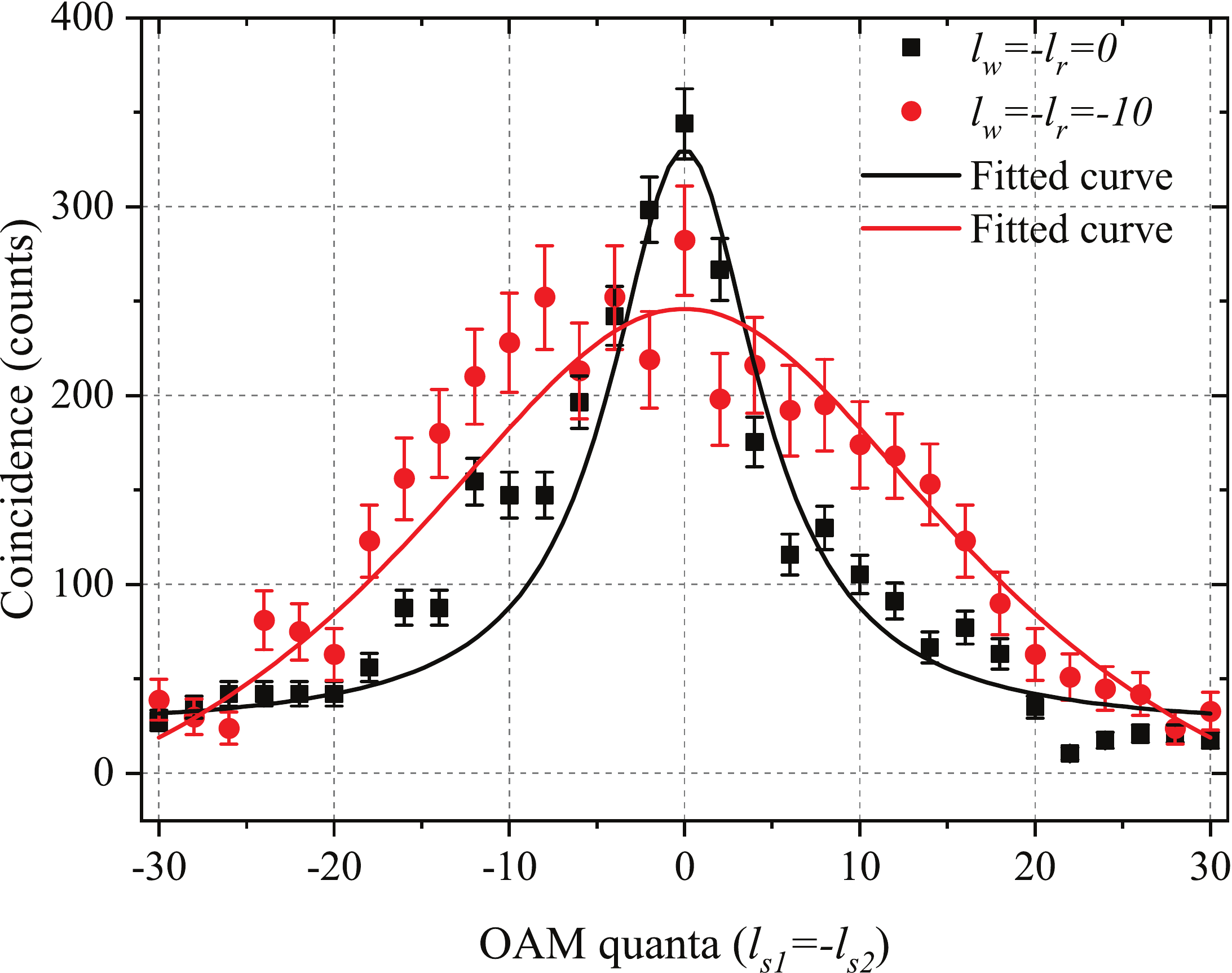}\caption{The measured correlated OAM distribution under different OAM modes
of write- and read-beams. The correlated OAM distribution with $l_{w}=-l_{r}=0$
(black line) and $l_{w}=-l_{r}=10$ (red line). These data are fitted
by the function $y={y_{0}}+2aw/[\pi(4{x^{2}}+{w^{2}})]$ with parameters
($y_{0}=15.8$, $w=11.4\pm2.3$, $a=5501$) and ($y_{0}=0,w=27.0\pm1.8,a=11534$). }

\label{result 2}
\end{figure}
In previous work \citep{ding2016high} for demonstrating high-D OAM
quantum interface with Gaussian mode input, it is hard to generate
higher-D entanglement because the correlated coincidences of photons
decay very quickly against $l$. Here, we turn the OAM quanta of write-
and read-beams to modulate the light-atom interaction length. We input
the write-laser with OAM quanta of $l_{W}$. Due to the fact that
SRS process conserves angular momentum, we have created OAM entanglement
between Signal 1 and atomic memory, which can be specified by the
formula

\begin{equation}
\left|\psi\right\rangle _{_{photon-atom}}^{l_{W}}{\rm =}\sum\nolimits _{l=-\infty}^{l=\infty}c_{l}\left|l\right\rangle _{{\rm S1}}\otimes\left|l_{W}-l\right\rangle _{a}
\end{equation}
here, $\left|c_{l}\right|^{2}$ represents excitation probability,
$\left|l\right\rangle _{{\rm S1}}$ is the OAM eigenmode of Signal
1 with quanta of $l$. $\left|l_{W}-l\right\rangle _{a}$ is the OAM
eigenmode of atomic memory with quanta of $l_{W}-l$. Through this
method, the atomic memory could carry the arbitrary OAM topological
charge with the term of $l_{W}-l$, thus resulting in the redistributed
quantum interface. We also check the conservation of OAM modes at
two situations $l_{W}=2$, $l_{R}=0$ and $l_{W}=1$ and $l_{R}=2$,
which can be found in the supplement. Most importantly, the spiral-bandwidth
of generated photons is broadened when we increase the OAM quanta
of write- and read-beams.

\begin{figure*}[t]
\includegraphics[width=1.8\columnwidth]{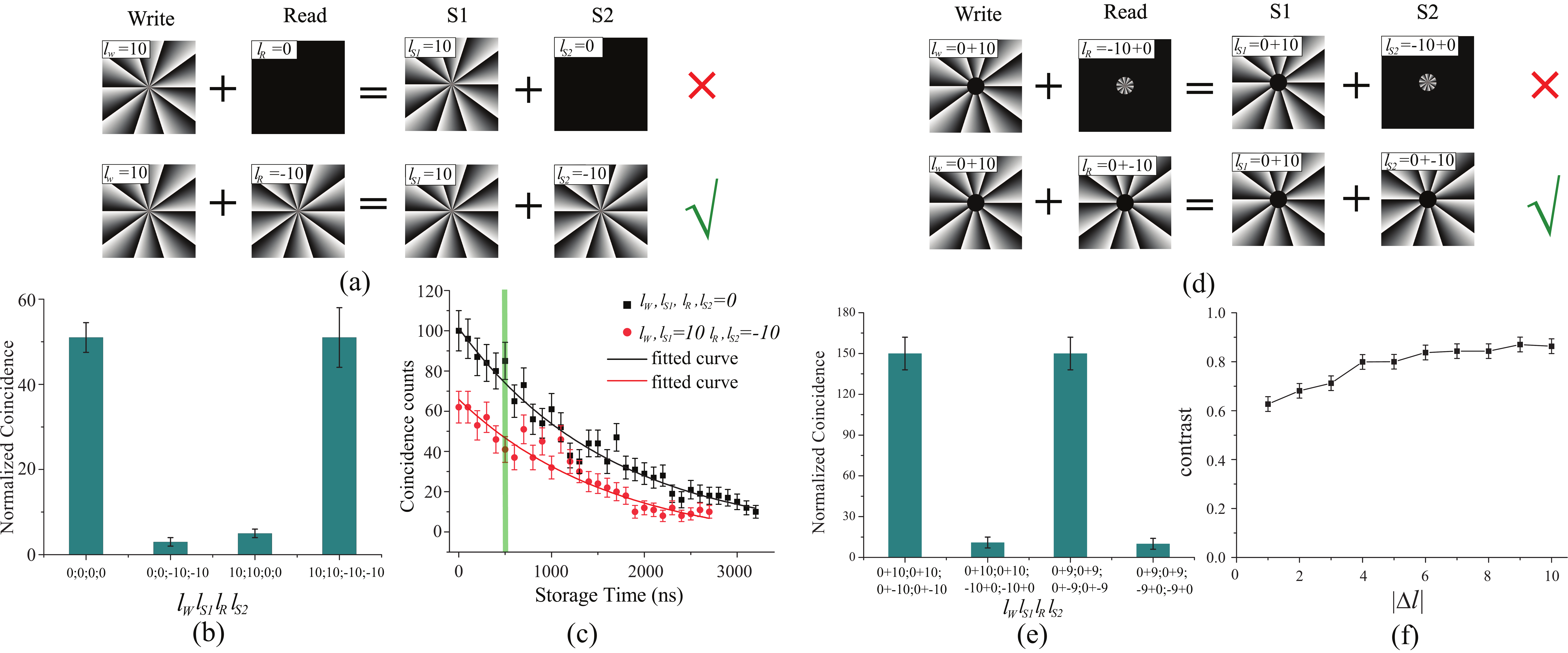}\caption{Multiplexing OAM modes. (a) The multiplexing with different OAM modes
of write- and read-lasers. The cross and check marks in up/down equation
shows the weak/strong correlation under the different ($|l_{W}|\protect\neq|l_{R}|$)
/same OAM orders ($|l_{W}|=|l_{R}|$). (b) The coincidence counts
for the different situations of $\lvert l_{W}\rangle=\lvert0\rangle/\lvert10\rangle$,
$\lvert l_{S1}\rangle=\lvert0\rangle/\lvert10\rangle$, $\lvert l_{R}\rangle=\lvert0\rangle/\lvert-10\rangle$,
$\lvert l_{S2}\rangle=\lvert0\rangle/\lvert-10\rangle$. (c) The memory
decay function with different OAM modes (black for $|l_{W}|=0$, red
for $|l_{W}|=10$). The memory decays exponentially with coherence
time of $\tau=1655$ ns. (d) The multiplexing in radial direction
with different OAM modes in inner and outer rings. When the OAM orders
in the inner and outer rings are different (up equation), the correlation
is weak; On the contrary, the correlation is strong when the OAM orders
are same (down equation). (e) The normalized coincidence counts with
$|l_{W}\rangle=|0\rangle+|10\rangle$, $|l_{R}\rangle=|-10\rangle+|0\rangle$
and $|l_{W}\rangle=|0\rangle+|10\rangle$, $|l_{R}\rangle=|0\rangle+|-10\rangle$
in the left two bars. The right two bars are the corresponding OAM
mode of $|l_{W}\rangle=|0\rangle+|9\rangle$, $|l_{R}\rangle=|-9\rangle+|0\rangle$
and $|l_{W}\rangle=|0\rangle+|9\rangle$, $|l_{R}\rangle=|0\rangle+|-9\rangle$.
(f) The detected contrast of coincidence counts against different
$\Delta l$. The storage time is set to be 500 ns. The different signs
of OAM quanta set in (a) and (d) are required to maintain OAM conservation.}

\label{result 3}
\end{figure*}
Due to the broadening effect of spiral-bandwidth with larger $l$
laser beam input, the distribution of generated OAM signal 1 and memory
would be redistributed in more flat range. This is because the generated
OAM modes are dependent on the interaction length and the waist of
the write- and read-beams \citep{du2008narrowband}. The vector mismatching
$\Delta k$ from transverse azimuthal phase would increase the value
of $\Delta k\cdot L$, where $L$ is the interaction length. This
effect is very promising because it is regarded as a concentration
operation. In order to achieve high-D OAM quantum memory, we utilize
the above method to extend the quanta of write-laser, we set $l_{W}=10$.
In addition, we set $l_{R}=-10$ for reading process. The writing
and reading process of DLCZ quantum memory is essentially a delayed
spontaneously four-wave mixing process. Based on the unique advantages
of individually modulating write- and read-beams of four-wave mixing
process (not like a single pump field used in spontaneous parametric
down conversion), the write- and read-laser beams can be individually
loaded OAM modes with opposite signs whilst the input total angular
momentum can be zero, thus making the joint spectrum of correlation
broadeded. We map different OAM phases onto SLM 2 and SLM 4, and record
the coincidence between Signal 1 and Signal 2 photons. The spiral-bandwidth
of OAM entanglement is measured in the red line in Fig. \ref{result 2}.
The spiral-bandwidth of generated single photons becomes much flat
than the scheme with inputting Gaussian mode. The spiral-bandwidth
of Gaussian mode is about $\delta l=11.4\pm2.3$, whilst for $l_{W}=-l_{R}=10$
the spiral-bandwidth is $\delta l=27.0\pm1.8$ obviously enhanced
by a factor of $\sim2.4$. We also check the high-D OAM entanglement
with the broadened spiral-bandwidth, and give a high-D entanglement
properties with OAM quanta up to $\Delta l=16$, see supplement.

With increasing the OAM quanta in DLCZ writing and reading processes,
we exhibit a potential application for multiplexing with different
OAM modes. If we select $l_{W}=10$ for writing and $l_{R}=-10$ for
reading out, we can detect the correlated coincidence; while for $l_{W}=10$,
$l_{R}=0$ given in Fig. \ref{result 3}(a), there would be almost
no coincidence counts exhibiting orthogonality-like property shown
by Fig. \ref{result 3}(b). The nonlinearity of the interleaved OAM
modes is strongly dependent on the overlap of beam profiles of write-
and read-beams, it would become small if the mismatch between write-
and read-beams is large, then the coincidence counts would jump down,
see supplement. The detected contrast $(C_{\text{same}}-C_{\text{diff}})/(C_{\text{same}}+C_{\text{diff}})$
is $0.85\pm0.03$ (where $C_{\text{same}}$ and $C_{\text{diff}}$
are defined as coincidence counts with same ($|l_{W}|=|l_{R}|$) and
different ($|l_{W}|\neq|l_{R}|$) write- and read-beam OAM modes),
near to the theoretical estimation of $0.91$. In this process, the
storage coherence time is almost same for $l_{W}=0$ and $l_{W}=10$,
see Fig. \ref{result 3}(c). Furthermore, we map the different OAM
modes in the inner ($l_{in}$) and outer rings ($l_{out}$) to detect
the multiplexing property along radial direction, see Fig. \ref{result 3}(d).
Since the nonlinearity of interleaved OAM modes (for example $l_{W}=0$
and $l_{W}=10$) in the center of ensemble can be distinguished by
inputting distinct OAM modes (Fig. \ref{result 3}(e)), this may result
in multiplexing along radial direction. We map the different OAM modes
with $\Delta l=1,2\ldots10$ in inner and outer rings ($\Delta l=l_{out}-l_{in}$)
and detect the correlation given in Fig. \ref{result 3}(e). The crosstalk
between different OAM modes is detected by setting same phase structure
or the different phase structure. The contrast of coincidence counts
is increased against with $\Delta l$ because the large difference
$\Delta l$ means that the mismatch between write- and read-beams
decrease the correlation (Fig. \ref{result 3}(f)), agreeing with
the above analysis. This whole process could be regarded as the multiplexing
of two OAM spectra, which are created by inputting two distinct OAM
writing and reading ($l_{out}$, $l_{in}$), thus satisfying some
quantum information protocols with one OAM spectra docking to another
OAM spectra.

\begin{figure}[t]
\includegraphics[width=1\columnwidth]{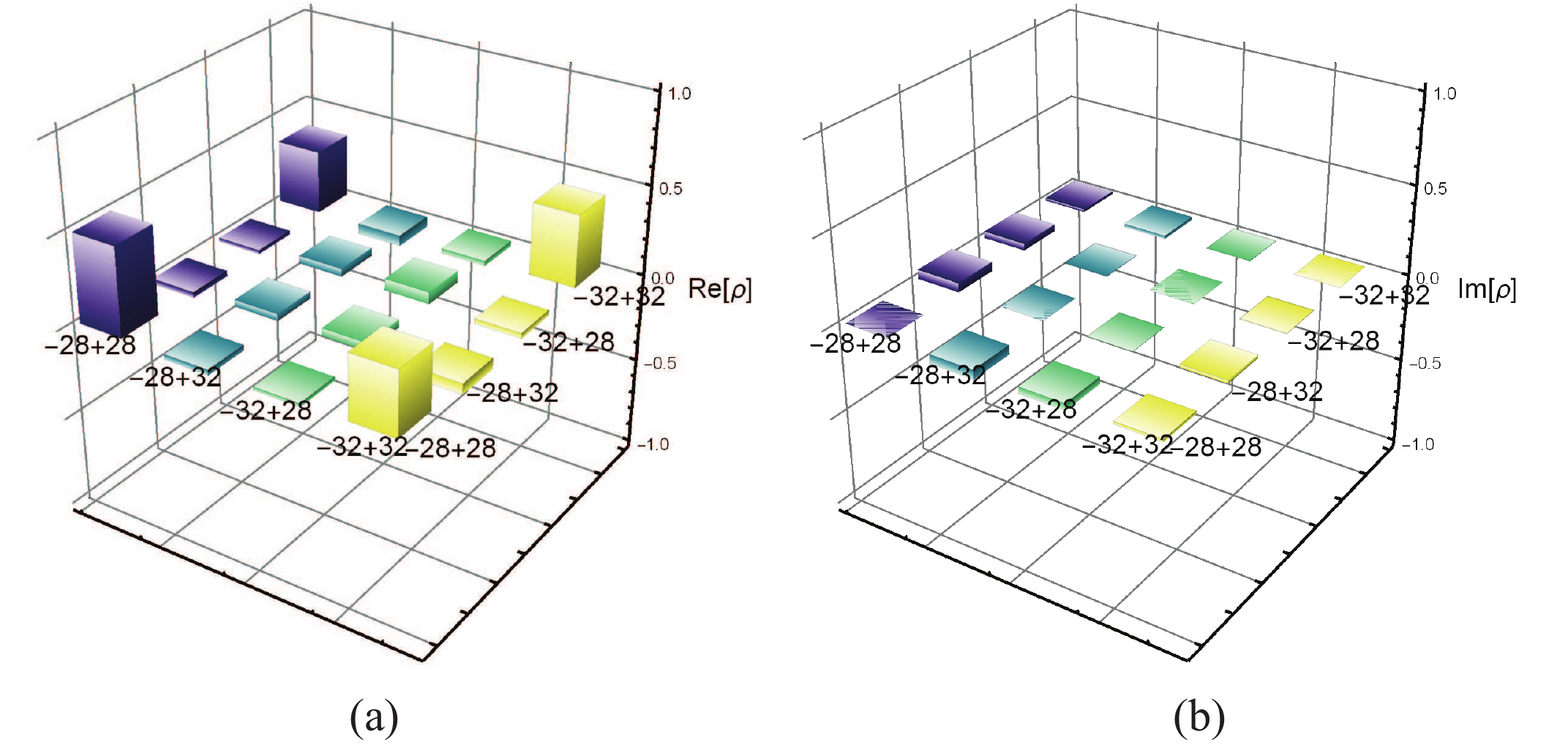}\caption{The reconstructed density matrix for photonic entangled state $\left|\psi\right\rangle _{photon-photon}^{{\rm {30}},{\rm {-30}}}$.
(a) and (b) are the real and imaginary parts of density matrix respectively.
Each data for reconstructing density matrices are recorded in 3000
s}

\label{result 4}
\end{figure}

At last, broad spiral-bandwidth offers an ability for demonstrating
high-quanta OAM quantum interface in 2-D subspace. For this, we set
$l_{W}=30$, $l_{R}=-30$, the storage time is set to be twice of
the width of write pulse, the decoherence from the transverse azimuthal
momentum mismatch between write-laser and the Signal 1 photons is
ignored. The nonlinearity of the DLCZ process at large quanta $l$
is relatively small, we then only consider the two OAM modes for verifying
entanglement. We choose the OAM modes of $l=28,32$ to verify the
high-quanta OAM entanglement. The photonic entangled state is expressed
as:

\begin{equation}
\left|\psi\right\rangle _{photon-photon}^{{\rm {30}},{\rm {-30}}}{\rm {=}}{\raise0.5ex\hbox{\ensuremath{{\scriptstyle {\rm {1}}}}}\kern-0.1em /\kern-0.15em \lower0.25ex\hbox{\ensuremath{{\scriptstyle {\sqrt{2}}}}}}\left({{{\left|{-28}\right\rangle }_{{\rm {S1}}}}{{\left|{28}\right\rangle }_{{\rm {S2}}}}+{{\left|{-32}\right\rangle }_{{\rm {S1}}}}{{\left|{32}\right\rangle }_{{\rm {S2}}}}}\right)
\end{equation}
Through quantum state tomography, we obtain the reconstructed density
matrix as shown in Fig. \ref{result 4}(a) and (b). The fidelity of
reconstructed density matrix is calculated as $80.5\pm4.8\%$ by comparing
with the ideal density matrix. We also check the violation of Clauser-Horne-Shimony-Holt
(CHSH) inequality \citep{clauser1969proposed,freedman1972experimental,clauser1974experimental}
to demonstrate the nonlocality of the entangled state. The CHSH parameter
S \citep{leach2009violation} is represented as following: $S=\left|E\left(\theta_{S2},\theta_{S1}\right)-E\left(\theta_{S2},\theta'_{S1}\right){\rm +}E\left(\theta'_{S2},\theta_{S1}\right){\rm +}E\left(\theta'_{S2},\theta'_{S1}\right)\right|$.
Here, the correlation function $E\left({{\theta_{S2}},{\theta_{S1}}}\right)$
can be calculated from the rates of coincidence at several particular
orientations, $\theta/\theta'$ represents the angle of phase distribution
on the surface of SLM 2 and SLM 4. The calculated S is $2.22\pm0.07$
which is larger than 2 violating the CHSH inequality, thus it demonstrates
the real entanglement between Signal 1 and Signal 2 photons.

In this work, we demonstrate a broad spiral-bandwidth OAM DLCZ memory,
the OAM distribution and the quanta of OAM quantum interface are freely
manipulated. In this state-of-the-art quantum interface, we have achieved
high-D OAM entanglement with OAM modes difference $\Delta l$ up to
16, the quanta of OAM 2-D subspace can be accessible to $l=\pm30$.
The experiment reported here would be very promising to demonstrate
high-quanta OAM quantum interface and study the fundamental physics
in OAM-based light and matter interaction.

\subsection*{ACKNOWLEDGMENTS}

Dong-Sheng Ding and Ming-Xin Dong contributed this paper equally.
We thank Guo-Yong Xiang professor for loaning a SLM. This work was
supported by National Key R\&D Program of China (2017YFA0304800),
the National Natural Science Foundation of China (Grant Nos. 61525504,
61722510, 61435011, 11174271, 61275115, 11604322), and the Innovation
Fund from Chinese Academy of Sciences.

\section*{Supplementary}

\subsection*{Experimental time sequence. }

The repetition rate of our experiment is $100\,\mathrm{Hz}$, and
the MOT trapping time is 8.7 ms. Besides, the operation window of
1.3 ms consists of 2600 cycles with a cycle time of 500 ns. Write-laser
and read-laser are pulsed by acousto-optic modulator with pulse width
of 50 ns and 200 ns respectively in each cycle. The optical depth
in MOT is about 40. The storage time is controlled by changing the
delay time between write- and read-laser through an arbitrary function
generator. The magnetic field for trapping is shut down in the experiment
window.

\subsection*{4-F image system for four SLMs. }

The SLM 1 acts as a mask plane, and the center of atomic ensemble
in MOT is the image plane. Two lenses L1 and L2 with focal length
of 300 mm and 500 mm are utilized to map the phase message of SLM
1 to the atomic ensemble. Due to the phase matching condition $k_{W}-k_{S1}=k_{R}-k_{S2}$,
the imaging system can be easily optically aligned. The Signal 1 and
Signal 2 fields are collinear, the Signal 1 beam is completely overlapped
by the write beam through demonstrating electromagnetically induced
transparency effect. Here, the write-laser carrying high OAM quanta
diffracts very strongly and results in the waist of laser beam too
large in the center of atomic ensemble, which results in weak interaction
between write-laser and atomic ensemble. Through the 4-f image system
with unequal arms, we can not only map the OAM phase message to the
center of atomic ensemble accurately but also decrease the waist of
write-laser with high OAM quanta. Similarly, the single photon carried
with OAM phase message from the center of atomic ensemble is retrieved
to project on SLM 1 via the other 4-f image system, and ultimately
we collect photons by single-mode fibers.

\subsection*{Theoretical analysis. }

In the interaction picture, despite the decay of spin wave, the effective
Hamiltonian for the delayed four-wave mixing process is written as
\citep{wen2006transverse}

\begin{align}
{\hat{H}_{I}} & =\frac{{\varepsilon_{0}}}{4}\int_{-L/2}^{L/2}{dz{\chi^{(3)}}{{\vec{E}}_{W}}{{\vec{E}}_{R}}{{\vec{E}}^{*}}_{S1}{{\vec{E}}^{*}}_{S2}}+H.c
\end{align}
where $H.c.$ means the Hermitian conjugate. ${\chi^{(3)}}$ is the
third-order nonlinear susceptibility for resonant signal 2 photon,
which is given \citep{braje2004frequency}:

\begin{align}
{\chi^{(3)}} & =\frac{{N{\mu_{13}}{\mu_{32}}{\mu_{24}}{\mu_{41}}/({\varepsilon_{0}}{\hbar^{3}})}}{{({\Delta_{W}}+i{\gamma_{23}})[{{\left|{\Omega_{R}}\right|}^{2}}-4(\omega+i{\gamma_{24}})(\omega+i{\gamma_{21}})]}}
\end{align}
here, ${\mu_{ij}}$ are the electric dipole matrix elements. ${\gamma_{ij}}$
are the dephasing rates. ${\Omega_{R}}$ is the Rabi frequency of
read laser. The probability to generate the Signal 1 and Signal 2
in modes $|l_{S1}\rangle$, $|l_{S2}\rangle$ is given by the overlap
with write- and read-laser beam profiles:

\begin{equation}
\begin{split}c_{l_{W}l_{R}l_{S1}l_{S2}} & \sim\int_{-L/2}^{L/2}\int_{0}^{r}\int_{0}^{2\pi}\varepsilon_{0}\chi^{(3)}rLG_{0}^{l_{W}}(r,\phi)LG_{0}^{l_{R}}(r,\phi)\\
 & LG_{0}^{l_{W}}(r,\phi)LG_{0}^{l_{R}}(r,\phi)[LG_{0}^{l_{S1}}(r,\phi)]^{*}[LG_{0}^{l_{S2}}(r,\phi)]^{*}d\phi drdz
\end{split}
\end{equation}
The integral over the azimuthal coordinate is

\begin{align}
\int_{0}^{2\pi}{d\phi}exp[i({l_{W}}+{l_{R}}-{l_{S1}}-{l_{S2}})\phi] & =2\pi{\delta_{{l_{W}}+{l_{R}},{l_{S1}}+{l_{S2}}}}
\end{align}
From which, we can obtain the topological charge conservation law
in OAM space is ${l_{W}}+{l_{R}}={l_{S1}}+{l_{S2}}$. According to
Eq. (4), we can find the probability of ${l_{S1}}$-Signal 1 and ${l_{S2}}$-Signal
2 photons with ${l_{W}}$-write and ${l_{R}}$-read lasers, which
strongly depends on the profiles match between the four fields.

In order to illustrate the topological charge conservation law in
our OAM quantum interface in DLCZ memory, we input the write-laser
with OAM quanta of $l_{W}$. Due to the fact that SRS process conserves
angular momentum, we have created OAM entanglement between Signal
1 and atomic spin wave, which can be specified by the formula $\left|\psi\right\rangle _{_{photon-atom}}^{{l_{W}}}{\rm {=}}\sum\nolimits _{l=-\infty}^{l=\infty}{c_{l}}{\left|l\right\rangle _{{\rm {S1}}}}\otimes{\left|{{l_{W}}-l}\right\rangle _{a}}$,
here, ${\left|{c_{l}}\right|^{2}}$ represents excitation probability,
${\left|l\right\rangle _{{\rm {S1}}}}$ is the OAM eigenmode of Signal
1 with quanta of $l$. ${\left|{{l_{W}}-l}\right\rangle _{a}}$ is
the OAM eigenmode of atomic spin wave with quanta of ${l_{W}}-l$.
Through this method, the atomic spin wave could carry the arbitrary
OAM topological charge with the term of ${l_{W}}-l$, thus resulting
in the redistributed quantum interface.

\begin{figure}[H]
\includegraphics[width=1\columnwidth]{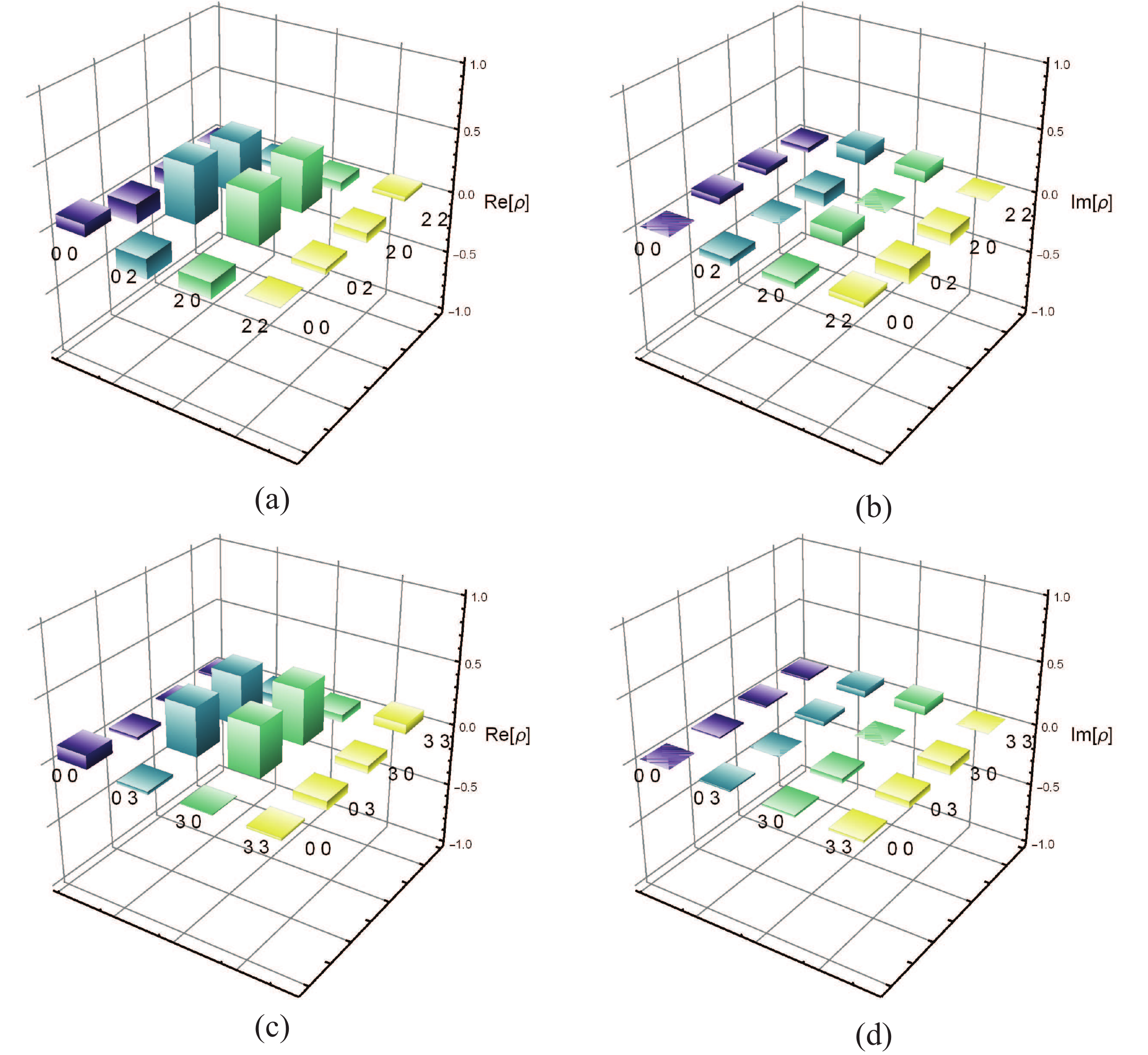}\caption{Reconstructed density matrices for Modulated OAM entanglement. The
real (a,c) and imaginary (b, d) parts of density matrices for photonic
OAM entangled state $\left|\psi\right\rangle _{photon-photon}^{2,0}$
and $\left|\psi\right\rangle _{photon-photon}^{1,2}$. Each data for
reconstructing density matrices are recorded in 1000 s.}

\label{result 1}
\end{figure}
After a period of storage, we check photon-atom entanglement by inputting
read-laser with OAM quanta of $l_{R}$, and checking the entanglement
between Signal 1 and Signal 2. The entanglement between Signal 1 and
Signal 2 can be written as $\left|\psi\right\rangle _{photon-photon}^{{l_{W}},{l_{R}}}{\rm {=}}\sum\nolimits _{l=-\infty}^{l=\infty}{c_{l}}{\left|l\right\rangle _{{\rm {S1}}}}\otimes{\left|{{l_{W}}+{l_{R}}-l}\right\rangle _{{\rm {S2}}}}$.
At first, we set $l_{W}=2$ and $l_{R}=0$, it means using OAM quanta
of 2 and 0 to write and read respectively. Thus, the photonic entangled
state is a sum of ${\left|l\right\rangle _{{\rm {S1}}}}\otimes{\left|{2-l}\right\rangle _{{\rm {S2}}}}$
with different $l$, this is a modulated asymmetric OAM entangled
state. Here, we only post-select the OAM mode of entangled state into
two-dimensional subspace ${\left|0\right\rangle _{{\rm {S1}}}}{\left|2\right\rangle _{S2}}$
and ${\left|2\right\rangle _{{\rm {S1}}}}{\left|0\right\rangle _{S2}}$,
that is $\left|\psi\right\rangle _{photon-photon}^{2,0}{\rm {=}}{\raise0.5ex\hbox{\ensuremath{{\scriptstyle 1}}}\kern-0.1em /\kern-0.15em \lower0.25ex\hbox{\ensuremath{{\scriptstyle {\sqrt{2}}}}}}\left({{{\left|0\right\rangle }_{{\rm {S1}}}}{{\left|2\right\rangle }_{{\rm {S2}}}}+{{\left|2\right\rangle }_{{\rm {S1}}}}{{\left|0\right\rangle }_{{\rm {S2}}}}}\right)$.
To characterize the OAM entanglement between Signal 1 and Signal 2,
we reconstruct the density matrices by projecting Signal 1 and Signal
2 onto OAM bases of $\left|0\right\rangle $, $\left|2\right\rangle $,
${{\left({\left|0\right\rangle -i\left|2\right\rangle }\right)}\mathord{\left/{\vphantom{{\left({\left|0\right\rangle -i\left|2\right\rangle }\right)}{2^{1/2}}}}\right.\kern-\nulldelimiterspace}{2^{1/2}}}$,
${{\left({\left|0\right\rangle +\left|2\right\rangle }\right)}\mathord{\left/{\vphantom{{\left({\left|0\right\rangle +\left|2\right\rangle }\right)}{2^{1/2}}}}\right.\kern-\nulldelimiterspace}{2^{1/2}}}$
for demonstrating quantum tomography. Then we use the obtained 16
coincidence rates to reconstruct the density matrix of state as shown
in Fig. \ref{result 1} (a) and (b). According to the formula $F{\rm {=Tr(}}\sqrt{\sqrt{\rho}{\rho_{{\rm {ideal}}}}\sqrt{\rho}}{{\rm {)}}^{2}}$,
which compares the constructed density matrix $\rho$ with the ideal
density matrix ${\rho_{{\rm {ideal}}}}$, we obtain the fidelity of
$83.3\pm3.5$\%. We also try another data set of $m_{1}=1$ and $m_{2}=2$,
and obtain the photonic entangled state $\left|\psi\right\rangle _{photon-photon}^{1,2}{\rm {=}}{\raise0.5ex\hbox{\ensuremath{{\scriptstyle 1}}}\kern-0.1em /\kern-0.15em \lower0.25ex\hbox{\ensuremath{{\scriptstyle {\sqrt{2}}}}}}\left({{{\left|0\right\rangle }_{{\rm {S1}}}}{{\left|3\right\rangle }_{{\rm {S2}}}}+{{\left|3\right\rangle }_{{\rm {S1}}}}{{\left|0\right\rangle }_{{\rm {S2}}}}}\right)$.
Similarly, we reconstruct the density matrix of this state, the real
and imaginary parts of reconstructed density matrix are shown in Fig.
\ref{result 1} (c) and (d), with fidelity of $81.1\pm4.2$\%. In
this process, although the fidelity is not very high, but it reveals
that in DLCZ quantum memory, the OAM modes are conserved in the whole
writing and reading process.

\subsection*{The entanglement dimensionality witness. }

In order to demonstrate the high-D entanglement between Signal 1 and
atomic memory, we avoid the crosstalk between neighboring OAM modes.
We select the modes of $l=0,4,8,12,16$ in which three modes between
adjacent terms are removed for better isolation. We read the photon-atom
entanglement out to photon-photon entanglement for verification. So,
the entangled state is $\left|\psi\right\rangle _{photon-photon}^{{\rm {10}},-{\rm {1}}0}{\rm {=}}{c_{1}}{\left|0\right\rangle _{{\rm {S1}}}}{\left|0\right\rangle _{{\rm {S2}}}}+{c_{2}}{\left|{-4}\right\rangle _{{\rm {S1}}}}{\left|4\right\rangle _{{\rm {S2}}}}+{c_{3}}{\left|{-8}\right\rangle _{{\rm {S1}}}}{\left|8\right\rangle _{{\rm {S2}}}}+{c_{4}}{\left|{-12}\right\rangle _{{\rm {S1}}}}{\left|{12}\right\rangle _{{\rm {S2}}}}+{c_{5}}{\left|{-16}\right\rangle _{{\rm {S1}}}}{\left|{16}\right\rangle _{{\rm {S2}}}}$here,
${c_{5}}\sim{c_{5}}$ are the corresponding amplitudes of different
terms ${\left|0\right\rangle _{{\rm {S1}}}}{\left|0\right\rangle _{{\rm {S2}}}}\sim{\left|{\rm {-16}}\right\rangle _{{\rm {S1}}}}{\left|{\rm {16}}\right\rangle _{{\rm {S2}}}}$.
For verifying the high-D state, it is very promising to use high-D
entanglement dimensionality witness \citet{agnew2012observation,krenn2014generation}
to characterize the entanglement existed in our system. The entanglement
dimensionality witness is expressed as ${W_{d}}=3\frac{{D(D-1)}}{2}-D(D-d)$,
here, $D$ is the number of measured OAM modes, and $d$ is associated
with dimensions of entanglement. If $W>{W_{d}}$, the two photons
entangled in at least $d+1$ dimensions, where $W$ is obtained from
calculating the sum of visibilities $N=V_{x}+V_{y}+V_{z}$ in total
two dimensional subspace. The $V_{x}$, $V_{y}$ and $V_{z}$ represent
the visibilities of two-photon interference in the diagonal/anti-diagonal,
left-circular/right-circular and horizontal/vertical bases respectively
in each OAM mode of $a$ and $b$, here $a$ and $b$ are selected
from $l=0,4,8,12,16$. A disadvantage of quantum tomography for high-D
entanglement is that the needed measurement data is the order of $d^{4}$,
which is a large challenge in its realization and is impractical while
$d$ is set to 5 in our experiment. Therefore, we adopt the method
of dimensionality witness to certificate the existence of high-D entanglement
and characterize the dimensionality. We calculate the value $W$ is
21.93\textpm 0.55, which violates the bound $W_{d}$ of 20 (the number
of measured OAM mode $D$ is 5 and $d$ is 3) by 3 s.d\textquoteright s,
thus there is at least a 4-D OAM entanglement between Signal 1 and
Signal 2 photons. In these measurements, the atom-photon entangled
states are both detected in photonic regime, we assume the fidelity
of reading out from ensembles is near unit. Although there are definitely
some noise or inefficient elements from reading process, making the
degree of the measured entanglement lower than that existed in ensembles.

\begin{figure}[H]
\includegraphics[width=1\columnwidth]{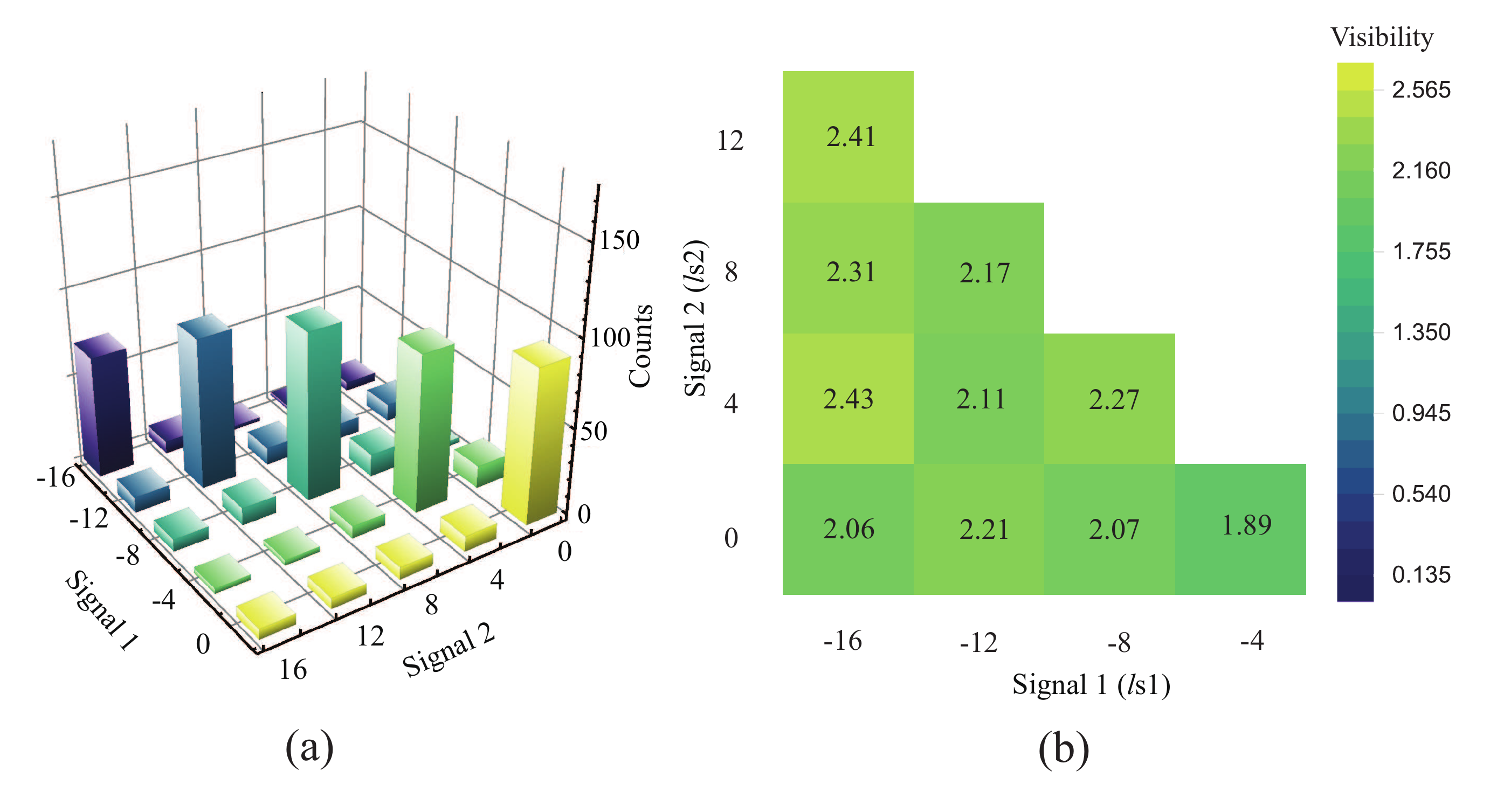}\caption{(a) The post-selected correlated OAM matrix between Signal 1 and Signal
2 photons with OAM modes difference $|\Delta l|$ up to 16. (b) The
each sum of visibilities for 2-D subspaces for detecting the high-D
entanglement dimensionality witness.}

\label{high-D}
\end{figure}

\subsection*{2-D high-$l$ Entanglement and state tomography}

If we considered the OAM modes of $a$ and $b$ with $l$=32 and 28,
the Signal 1 and Signal 2 are entangled in OAM space and entangled
state is expressed as
\begin{equation}
\left|{\Psi_{3}}\right\rangle {\rm {=}}{\raise0.5ex\hbox{\ensuremath{{\scriptstyle {\rm {1}}}}}\kern-0.1em /\kern-0.15em \lower0.25ex\hbox{\ensuremath{{\scriptstyle {\sqrt{{\rm {2}}}}}}}}\left({{{\left|{-28}\right\rangle }_{{\rm {S1}}}}{{\left|{28}\right\rangle }_{{\rm {S2}}}}+{{\left|{-32}\right\rangle }_{{\rm {S1}}}}{{\left|{32}\right\rangle }_{{\rm {S2}}}}}\right)
\end{equation}
Here, ${\left|{-28}\right\rangle _{{\rm {S1}}}}$ represents the Signal
1 carrying with OAM quanta of -28. By using two computers, we project
two photons onto two SLMs respectively and four state of $\left|{\phi_{1\sim4}}\right\rangle $
($\left|{-28}\right\rangle $, $\left|{-32}\right\rangle $, ${{\left({\left|{-28}\right\rangle -i\left|{-32}\right\rangle }\right)}\mathord{\left/{\vphantom{{\left({\left|{-28}\right\rangle -i\left|{-32}\right\rangle }\right)}{2^{1/2}}}}\right.\kern-\nulldelimiterspace}{2^{1/2}}}$,
${{\left({\left|{-28}\right\rangle +\left|{-32}\right\rangle }\right)}\mathord{\left/{\vphantom{{\left({\left|{-28}\right\rangle +\left|{-32}\right\rangle }\right)}{2^{1/2}}}}\right.\kern-\nulldelimiterspace}{2^{1/2}}}$)
are programmed onto SLM 2 and $\left|{\varphi_{1\sim4}}\right\rangle $
($\left|{28}\right\rangle $, $\left|{32}\right\rangle $, ${{\left({\left|{28}\right\rangle -i\left|{32}\right\rangle }\right)}\mathord{\left/{\vphantom{{\left({\left|{28}\right\rangle -i\left|{32}\right\rangle }\right)}{2^{1/2}}}}\right.\kern-\nulldelimiterspace}{2^{1/2}}}$,
${{\left({\left|{28}\right\rangle +\left|{32}\right\rangle }\right)}\mathord{\left/{\vphantom{{\left({\left|{28}\right\rangle +\left|{32}\right\rangle }\right)}{2^{1/2}}}}\right.\kern-\nulldelimiterspace}{2^{1/2}}}$)
are programmed onto SLM 4. Then, we obtain a set of 16 data for reconstructing
the density matrix given in the main text. The error bars in our experiment
are estimated by Poisson statistics and using Monte Carlo simulations
with the aid of Mathematica software.

\end{document}